\documentclass[11pt]{article}

\usepackage{graphicx}

\begin{document}

{\bf
Halley's comet of 87 BC on the coins of Armenian king Tigranes? 
}

\vspace{0.1in}

\begin{center}

\noindent V.G.Gurzadyan$^{1}$ and R.Vardanyan$^2$

\end{center}

\vspace{0.05in}

$^1$ Yerevan Physics Institute, Armenia and University "La Sapienza", Rome, Italy; E-mail: gurzadya@icra.it

$^2$ History Museum, Yerevan, Armenia

\vspace{0.1in}

{\it (Published in "Astronomy \& Geophysics" (RAS, London), vol.45, p.4.6, 2004.)} 

\vspace{0.2in}

Coins of Armenian king Tigranes the Great clearly reveal a star with a tail on the royal tiara which may be associated with the Halley's passage of 87 BC.

Tigranes II the Great (95-55 BC) had made Armenia one of most powerful kingdoms in Western Asia, extending from the Caucasus to the eastern Mediterranean.  Economic needs associated with the expansion of the empire resulted in diverse silver and copper-bronze coinage. Although the chronological and mint problems of these abundant emissions of Tigranes's coins have been discussed by scholars, many questions remain. 

On many coins, notably the tetradrachms struck in Antioch in Syria, as well as the silver and bronze coins struck in Artaxata or Tigranocerta, Tigranes wears a tiara decorated with a sun/star symbol between two eagles.  On the tetradrachms and copper-bronze coins struck in Damascus, the king's tiara is adorned with a big sun/star symbol on the left side along with an eagle-like symbol to the right.  

On a rare series of tetradrachms and drachms and on more numerous copper coins (Figs.1,2) depicting the goddesses Tyche and Nike, cypress tree, palm branch, tripod on their reverse, Tigranes's tiara is decorated with a single star, which has one of the righthand-side rays elongated and  curved, which can be interpreted as a comet (Bedoukian 1978, Nercessian 2000a and b).  The king also looks younger on these coins. 

This latter series is the most mysterious in the entire Tigranes's coins regarding its dating and place of minting, as well as for the interpretation of the comet symbol.

All Tigranes's silver coins (except for the Damascus series and a large group of bronze coins) have on their reverse a goddess sitting on a rock, wearing a turreted crown, holding a palm branch and having a swimming nude figure of a river-god at her feet.  

Scholars have no doubts that this represents the bronze statue of the city goddess Tyche, created by Eutychides, the disciple of Lissipe, and erected in Antioch at the time of Seleucus I (312-280 BC).  Tigranes conquered this city in 83 BC.  This fact is commonly agreed, as it is based on a passage from Appian, as well as on the coins struck until this year in Antioch by the Seleucid king Antiochus X. Hence, the series with comet must have been struck after 83 BC.

Among the astronomical records of the period corresponding to the reign of Tigranes is the appearance of Halley's comet. According to the chronicles and modern backwards orbit computations, Halley passed the perihelion on 6 August 87 BC, being discovered in July and was last seen on August 24 (Kronk 1999).
Chinese chronicles of Han Shu (see Kronk 1999) mention also a comet in 84 BC and three astronomical events in 69 BC: "a sparkling star" seen during 27 January - 24 February 69 BC (probably a nova), a guest star observed on July 22, and another guest star discovered on 20 August and last seen on 27 August. Another sparkling star is noted in 61 BC.       

If the image on Tigranes's tiara is a comet, and it has to be attributed to a specific comet, then the appearance of Halley within his reign has to be the most significant event. If so, this is another case when astronomical events can also be useful for historical problems (Gasche et al 1998, Gurzadyan 2000).

In addition, this would be a far earlier record of Halley in Armenia than was previously known from chronicles (Gurzadyan 1988a,b) and also one of the earliest known images of Halley's comet. The curved (not straight) tail clearly visible on the coins also can act as comet's distinguishing information.

We thank Philippe Amram for information and Aram Gurzadyan for processing the images. 

\vspace{0.2in}
References
\vspace{0.1in}

Bedoukian P.Z. 1978 {\it Coinage of the Artaxiads of Armenia [ =RNS Special Publication 10]}, London, 48 pl.1: 10-11.

Gasche H., Armstrong J.A., Cole S.W., Gurzadyan V.G. 1998 {\it Dating the Fall of Babylon. 
Reappraisal of Second-Millennium Chronology}, Chicago University Press.

Gurzadyan V.G. 1988a {\it The Observatory} 108 127.

Gurzadyan V.G. 1988b {\it QJRAS} 29 275.
 
Gurzadyan V.G. 2000 Astronomy and the Fall of Babylon, {\it Sky and Telescope} 100 40.

Kronk G.W. 1999 {\it Cometography. A Catalog of Comets}, vol.1, Cambridge University Press.

Nercessian Y.T. 2000a Silver Coins of Tigranes II of Armenia, {\it Armenian Numismatic Journal} XXVI 91-92 pl.7-8. 

Nercessian Y.T. 2000b Tigranes the Great of Armenia and the Mint of Damascus, in: Y.T. Nercessian, {\it Armenian Numismatic Studies} Los Angeles 95-107 pl.27.

\end{document}